\documentclass[12pt]{article}
\usepackage{latexsym}
\begin{document}

\title{A new picture on (3+1)D topological mass mechanism}
\author{ O.S. Ventura$^{a,b,c}$, R. L. P. G. Amaral$^d$, J V Costa$^d$,\\ L. O. Buffon$^{b,e}$ and V. E. R. Lemes$^f$
\footnote{ozemar@uvv.br, rubens@if.uff.br, costa@if.uff.br, lobuffon@terra.com.br, vitor@dft.if.uerj.br }\\
\small\em $^a$Coordenadoria de Matem\'{a}tica, Centro Federal de Educa\c {c}\~{a}o Tecnol\'{o}gica \  do Esp\'{\i}rito Santo,  \\
\small\em $ $Avenida Vit\'{o}ria 1729 - Jucutuquara, Vit\'{o}ria - ES, 29040 - 333, Brazil\\
\small\em $^b$Centro Universit\'{a}rio de Vila Velha, Rua comiss\'{a}rio Jos\'{e} \\
\small\em Dantas de Mello 15 - Boa Vista, Vila Velha - ES, 29102 - 770, Brazil\\
\small\em $^c$Uniest, Av. Brasil 110, Jardim Am\'erica, Cariacica-ES, 29140 - 490, Brazil\\
\small\em $^d$Instituto de F\'\i sica, Universidade Federal Fluminense, 24210 - 340, Niter\'oi - RJ, Brasil,\\
\small\em $^e$Escola Superior de Ci\^{e}ncias da Santa Casa de Miseric\'{o}rdia de Vit\'{o}ria,\\ \small\em Av. Nossa Senhora da Penha 2190, Santa Luiza, Vit\'{o}ria-ES, 29045-402, Brazil\\
\small\em $^f$Instituto de F\'\i sica, Universidade do Estado do Rio de Janeiro,\\ 
\small\em Rua S\~{a}o Francisco Xavier 524, Maracan\~{a}, Rio de Janeiro - RJ, 20550-013, Brazil.}
\date{\today}

\bigskip
\maketitle

\begin{abstract}

We present a class of mappings between the fields of the Cremmer-Sherk and pure
BF models in 4D. These mappings are established by two distinct procedures.
First a mapping of their actions is produced iteratively resulting in an
expansion of the fields of one model in terms of progressively higher derivatives 
of the other model fields. Secondly an exact mapping is introduced by mapping
their quantum correlation functions. The equivalence of both procedures is 
shown by resorting to the invariance under field scale transformations of 
the topological action. Related equivalences in 5D and 3D are discussed.
A cohomological argument is presented to provide consistency of the iterative
mapping.

\setcounter{page}{0}\thispagestyle{empty}
\end{abstract}

\section{Introduction}

The search for ultraviolet renormalizable models has always been one of the
most attractive and relevant aspects of quantum field theory. As it is well
known, the program of describing the eletro-weak interactions in the language
of QFT is based on the construction of the Higgs mechanism
for mass generation of the vector bosons. However, this mechanism relies on
the existence of a scalar particle, the Higgs boson, whose experimental
evidence is still lacking.

In this context, the topological mechanism for mass generation is atractive,
since it provides  masses for the gauge vector bosons without the explicit
introduction
of new scalar fields. For example, in three-dimensional spacetime, the
topological non-Abelian Chern-Simons term generates mass for the Yang-Mills
fields while preserving the exact gauge invariance $\cite{ng1}$. In four dimensions
the topological  mass generation mechanism occurs in the case of an anti-symmetric 
tensorial
field $B_{\mu \nu }$. It has been shown that the Cremmer-Sherk  action
gives a massive pole to the vector gauge field in the Abelian context. This
model is described by the action $\cite{ng2}$

\begin{equation}\label{cs}
S=\int d^4x\left( -\frac 14F^{\mu \nu }F_{\mu \nu }+\frac 1{12}H_{\mu \nu
\rho }H^{\mu \nu \rho }+\frac m4\varepsilon ^{\mu \nu \rho \sigma }F_{\mu
\nu }B_{\rho \sigma }\right) \,.  \label{bffgh}
\end{equation}
Indeed, as it was shown $\cite{nogo},$ this model exists only in the Abelian
version. In fact, possible non-Abelian generalizations of the action $\left( 
\ref{bffgh}\right) $, will necessarily require non-renormalizable couplings,
as in $\cite{ng4}$, or the introduction of extra fields $\cite{ng5}$.
Anti-symmetric fields in four dimensions also deserve attention since they
appear naturally by integrating out the fermionic degrees of freedom  in favor
of bosonic fields in bosonization approaches. The fermionic current turns
out to be expressed in terms of derivatives of the tensorial field as a topologically
conserved current. The coupling of this current 
to the gauge field leads to terms in the effective action similar to the
last one in (\ref{bffgh}).

An important property of the three dimensional Yang-Mills type actions, in
the presence of the Chern-Simons term, was pointed out in $\cite{simple}$,
i.e., it can be cast in the form of a pure Chern-Simons action through a
nonlinear covariant redefinition of the gauge connection. The quantum
consequences of this fact were investigated in the BRST framework yielding
an algebraic proof of the finiteness of the Yang-Mills action with
topological mass.

In this work we present a recursive mapping between Cremmer-Sherk's action
and the pure topological $BF$ model. With this the fields of one action are
expressed as a series of progressively higher derivatives of the other
model fields. This mapping is also established along a different line
in which the propagators of one action are reproduced using a closed
expression in terms of the other action fields.  This
 exposes the non-local nature of the mapping. 
 Related mappings  in higher and lower dimensions are discussed.
  A cohomological argument is presented to give consistency to the recursive mapping.

\section{Mapping the fields}

The aim of this section is to establish the classical equivalence between the
Cremmer-Sherk's action and the pure $BF\,$ theory, \textit{i. e.}, that the
first action can be mapped to the second one through a redefinition of
the gauge field. Following the same steps of the three-dimensional case $\cite
{simple}$, we search for a  redefinition 
 of the fields $A_\mu $ and $B_{\mu \nu }$ as a series in powers of $1/m$ in terms
of the fields   $\hat A_\mu $ and $\hat B_{\mu \nu }$ in
 such a way that the
relation below is valid\footnote{  We work in  the Minkovski spacetime so that 
$\varepsilon ^{\mu \nu \rho \sigma }\varepsilon _{\mu \nu \alpha \beta
}=-2\left( \delta _\alpha ^\rho \delta _\beta ^\sigma -\delta _\beta ^\rho
\delta _\alpha ^\sigma \right) 
$. We use $\varepsilon ^{0123}=1$ and $\mbox{diag}\;\eta_{\mu\nu}=(1,-1,-1,-1)$.
}: 
\begin{equation}
{\mathcal{S}_M}(A)\;+\;{\mathcal{S}_H}(B)+{\mathcal{S}_{BF}}(A,B)=%
{\mathcal{S}_{BF}}(\widehat{A},\widehat{B})\;,  \label{bff1}
\end{equation}
where 
\begin{eqnarray}\label{csandbf}
{\mathcal{S}_M}(A) &=&-\frac 14\int d^4x\left( F^{\mu \nu }F_{\mu \nu
}\right) \,, \\
{\mathcal{S}_H}(B) &=&+\frac 1{12}\int d^4x\left( H_{\mu \nu \rho
}H^{\mu \nu \rho }\right) \,, \\\label{purebf}
{\mathcal{S}_{BF}}(A,B) &=&\frac m4\int d^4x\left( \varepsilon ^{\mu \nu
\rho \sigma }F_{\mu \nu }B_{\rho \sigma }\right) \,
\end{eqnarray}
and the curvatures $F_{\mu \nu }$ and $H_{\mu \nu \rho }$ are the same given
in $\left( \ref{bffgh}\right) $, \textit{i. e.},

\[
F_{\mu \nu }=\partial _\mu A_\nu -\partial _\nu A_\mu 
\]
and 
\[
H_{\mu \nu \rho }=\partial _\mu B_{\nu \rho }+\partial _\rho B_{\mu \nu
}+\partial _\nu B_{\rho \mu }\,. 
\]
Indeed taking the field redefinitions in the form

\begin{eqnarray}\label{recursive1}
\widehat{A}_\mu &=&A_\mu +\sum_{n=1}^\infty \frac 1{(2m)^n}\vartheta _\mu ^n\,,
\label{bff3} \nonumber\\
\widehat{B}_{\mu \nu } &=&B_{\mu \nu }+\sum_{n=1}^\infty \frac 1{(2m)^n}\phi
_{\mu \nu }^n\,,  
\end{eqnarray}
the equality expressed in  $\left( \ref{bff1}\right) $ is implemented by 
recursively fixing the terms. We find  the expressions

\begin{eqnarray}\label{recursive2}
\phi_{\mu \nu }^{2n+1}&=&-\frac{ b_{2n+1}} {2}\epsilon_{\mu\nu\alpha\beta}\Box^n
F^{\alpha \beta}, \nonumber\\
\vartheta _\mu^{2n+1}&=&\frac {c_{2n+1}} {3} \epsilon_{\mu\nu\alpha\beta}\Box^n
H^{\nu \alpha\beta}, \nonumber\\
\phi_{\mu \nu }^{2n}&=&b_{2n}\Box^{n-1}
\partial^\alpha H_{\alpha\mu\nu}, \nonumber\\
\vartheta _\mu^{2n}&=&c_{2n} \Box^{n-1}
\partial^\nu F_{\nu\mu}, 
\end{eqnarray}
where the constants are defined as
\begin{eqnarray}
b_{2n+1}&=&-\sum_{j=1}^n c_{2j} b_{2(n-j)+1},\\
c_{2n+1}&=&-\sum_{j=1}^n b_{2j} c_{2(n-j)+1},\\
b_{2n}&=&B_n\left[2\sum_{j=1}^n b_{2j-1}c_{2(n-j)+1}-\sum_{j=1}^{n-1}
b_{2j}c_{2(n-j)}\right],\\
c_{2n}&=&(1-B_n)\left[2\sum_{j=1}^n b_{2j-1}c_{2(n-j)+1}+\sum_{j=1}^{n-1}
b_{2j}c_{2(n-j)}\right].
\end{eqnarray}
Here $B_n$ are arbitrary constants introduced at each even step of the process while
$b_1=-1$ and $c_1=-1/2$.

 The first terms can be expressed as

\begin{eqnarray}\label{bff6a}
\vartheta _\mu^{1}&=&\frac{-1}{6} \varepsilon^{\mu\nu\alpha\beta}
H_{\nu \alpha\beta }\nonumber,\\
\vartheta _\mu^{2}&=&(1-B_1) \partial_\nu F^{\nu\mu},\nonumber\\
\vartheta _\mu^{3}&=&\frac{B_1}{6} \varepsilon_{\mu\nu\alpha\beta}\Box
H^{\nu \alpha\beta}, \nonumber\\
\vartheta _\mu^{4}&=&(-1-B_1+B_1^2)(1-B_2) \Box
\partial_\nu F^{\nu\mu},\nonumber \\
\vartheta _\mu^{5}&=&\frac{-B_1^2-B_2-B_2B_1+B_2B_1^2}6 \varepsilon_{\mu\nu\alpha\beta}\Box^2
H^{\nu \alpha\beta },\nonumber\\
\vartheta _\mu^{6}&=&  \left[  B_1^2+B_2-B_2B_1-3B_2B_1^2+B_1+2-B_1^3+2B_1^3B_2  \right](1-B_3) \Box^{2}
\partial_\nu F^{\nu\mu}, \nonumber\\
\phi_{\mu \nu }^{1}&=&-\frac 12\varepsilon_{\mu\nu\alpha\beta}F^{\alpha \beta},\nonumber \\
\phi_{\mu \nu }^{2}&=&\frac{B_1}3\partial_\alpha H^{\alpha\mu\nu},\nonumber \\
\phi_{\mu \nu }^{3}&=&-\frac{1-B_1}{2}\varepsilon_{\mu\nu\alpha\beta}\Box 
F^{\alpha \beta},\nonumber \\
\phi_{\mu \nu }^{4}&=&\frac{B_2\left(-1-B_1+B_1^2\right)}3\Box
\partial_\alpha H^{\alpha\mu\nu}, \nonumber\\
\phi_{\mu \nu }^{5}&=&\frac{2-B_1-B_2-B_2B_1+B_2B_1^2  }2   
\varepsilon_{\mu\nu\alpha\beta}\Box^2 F^{\alpha \beta}\nonumber \\
\phi_{\mu \nu }^{6}&=&\frac{\left( B_1^2+B_2-B_2B_1-3B_2B_1^2+B_1+2-B_1^3+2B_1^3B_2\right) B_3}3
   \Box^{2}\partial_\alpha H^{\alpha\mu\nu}. 
\end{eqnarray}

As we can see, up to the sixth order in the mass parameter, the
coefficients, $\phi _{\mu \nu }^n\,$and $\vartheta _\mu ^n,$ shown in 
 (\ref{bff6a}), depend on the three arbitrary dimensionless parameters,
$B_1$, $B_2$ and $B_3$. In
fact at each new even  order in $1/m$ a new arbitrary parameter is allowed to be introduced.
As we shall see this is to be expected.

The formal series $\left( \ref{bff3}\right) ,$ which redefine the fields $%
A_\mu $ and $B_{\mu \nu },$ give the mapping we were looking for. 

Note that the gauge symmetry of the Cremmer-Sherk action is expressed as

\begin{equation}
\delta ^gA_\mu =\partial _\mu \varepsilon
,\,\,\,\,\,\,\,\,\,\,\,\,\,\,\,\delta ^gB_{\mu \nu }=0  \label{bhg}
\end{equation}
and 
\begin{equation}
\delta ^tA_\mu =0,\;\;\;\;\;\,\,\delta ^tB_{\mu \nu }=\partial _\mu
\varepsilon _\nu -\partial _\nu \varepsilon _\mu \,,  \label{mjk}
\end{equation}
\noindent
while the BF topological action is invariant under analogous transformations
\begin{equation}
\delta ^g\hat A_\mu =\partial _\mu \hat \varepsilon
,\,\,\,\,\,\,\,\,\,\,\,\,\,\,\,\delta ^g\hat B_{\mu \nu }=0  \label{bhg2}
\end{equation}
and 
\begin{equation}
\delta ^t\hat A_\mu =0,\;\;\;\;\;\,\,\delta ^t\hat B_{\mu \nu }=\partial _\mu
\varepsilon _\nu -\partial _\nu\hat \varepsilon _\mu \,.  \label{mjk2}
\end{equation}
The mapping (\ref{recursive1},\ref{recursive2})translates the gauge transformations of one pair of fields
straightforwardly into the ones of the other pair in such a way that $\hat \varepsilon$ and
$\hat \varepsilon_\mu$ are identified as  $ \varepsilon$ and
$ \varepsilon_\mu$. This occurs since the higher order terms in (\ref{recursive2}) are
gauge invariant. Indeed this result is natural but not mandatory in this procedure. The mappings
in (\ref{recursive1}) can be generalized by allowing for gauge dependence in the higher order terms.
 Due to the gauge symmetry such arbitrary terms would  not contribute to the action and would not spoil  the mapping.

\subsection{Exact mapping}

Let us express here the Cremmer-Sherk fields in terms of the pure BF 
fields using a new procedure. The Cremmer-Sherk propagators are given
by
\begin{eqnarray}\label{propcremmer}
<\imath {\cal{T}}B_{\mu\nu}B_{\alpha\beta}>&=&\left(P_{\mu\nu ,\alpha\beta}+G_1K_{\mu\nu,\alpha\beta}\right)\frac2{\Box(\Box+m^2)},\nonumber \\
<\imath {\cal{T}}B_{\mu\nu}A_{\alpha}>&=&-<\imath{\cal{T}}A_{\alpha}B_{\mu\nu}>=\left(S_{\mu\nu \alpha}\right)\frac m{\Box(\Box+m^2)},\nonumber \\
<\imath {\cal{T}}A_{\mu}A_{\alpha}>&=&\left(P_{\mu,\nu }+G_2K_{\mu,\nu}\right)\frac{-1}{\Box(\Box+m^2)}.
\end{eqnarray}

The pure BF propagators are  given by  

\begin{eqnarray}\label{propbf}
<\imath{\cal{T}}\widehat B_{\mu\nu}\widehat B_{\alpha\beta}>&=&\left(\widehat G_1K_{\mu\nu,\alpha\beta}\right)\frac2{m\Box}
,\nonumber\\
<\imath {\cal{T}}\widehat B_{\mu\nu}\widehat A_{\alpha}>&=&-<\imath {\cal{T}}\widehat A_{\alpha}\widehat B_{\mu\nu}>=\left(S_{\mu\nu \alpha}\right)\frac 1{m\Box},\nonumber \\
<\imath{\cal{T}}\widehat A_{\mu}\widehat A_{\alpha}>&=&\left(\widehat G_2K_{\mu,\nu}\right)\frac1{m\Box}.
\end{eqnarray}

Here the projectors are given by:

\begin{eqnarray}\label{projectors}
P_{\mu\nu,\alpha\beta}&=&\frac12\delta_{\mu\nu,\alpha\beta}\Box-\frac12K_{\mu\nu,\alpha\beta},\nonumber \\
K_{\mu\nu,\alpha\beta}&=&\delta_{\mu[\alpha}\partial_\nu\partial_{\beta ]}-\delta_{\nu[\alpha}\partial_\mu\partial_{\beta ]},\nonumber \\
S_{\mu\nu\alpha}&=&\varepsilon_{\mu\nu\alpha\beta}\partial^\beta ,\nonumber\\
P_{\mu\nu}&=&\delta_{\mu\nu}\Box-\partial_\mu\partial_\nu ,\nonumber\\
K_{\mu\nu}&=&\partial_\mu\partial_\nu.
\end{eqnarray}
The parameters $G$ and $\widehat G$ are introduced to fix the gauge.

Let us try to express the fields as

\begin{eqnarray}\label{mapping1}
A_\mu&=&\left(C_{AA}P_{\mu,\nu}+\delta_{\mu\nu}\right)\widehat A^\nu +C_{AB} S_{\mu\alpha\beta}\widehat B^{\alpha\beta},\nonumber\\
B_{\mu\nu}&=&C_{BA}S_{\mu\nu\alpha}\widehat A^\nu +\left(C_{BB} P_{\mu\nu,\alpha\beta}+\frac12\delta{\mu\nu,\alpha\beta}\right)
\widehat B^{\alpha\beta}.
 \end{eqnarray}
Computing the correlators of the Cremmer-Sherk field using this mapping and comparing with
(\ref{propcremmer}) the structure functions
are fixed. They result to be given by the non-local operators

\begin{eqnarray}\label{strucfunc}
C_{AA}&=&\frac{{2}^{\frac12}m^{\frac12}}{\sigma \Box }\left[\frac{
m-\sqrt{m^2+\Box}}{m^2+\Box}\right]^\frac12  -\frac1\Box ,\nonumber\\
C_{BB}&=&\frac{{\sigma}m^{\frac12}}{2^{\frac32}\Box }\left[\frac{
m-\sqrt{m^2+\Box}}{m^2+\Box}\right]^\frac12 -\frac1\Box ,\nonumber\\
C_{AB}&=&\frac{{\sigma}m^{\frac12}}{2^{\frac52}}\left[
\left(m-\sqrt{m^2+\Box}\right)\left(m^2+\Box\right)\right]^{-\frac12}  ,\nonumber\\
C_{BA}&=&\frac{{2}^{\frac12}m^{\frac12}}{\sigma}\left[
\left(m-\sqrt{m^2+\Box}\right)\left(m^2+\Box\right)\right]^{-\frac12}.  
\end{eqnarray}

Note that the non-local operators indeed map local fields of local models, 
the Cremmer-Sherk and pure BF models.
Observe   the presence of the arbitrary operator $\sigma$ in these equations. Its presence should be
expected since the set of transformations

\begin{eqnarray}
\hat A &\longrightarrow& \sigma \hat A,\nonumber \\
\hat B &\longrightarrow& \frac{1}{\sigma}\hat B,
\end{eqnarray}
does not affect any correlator of the BF model. The presence of  ${\sigma}$ in the
 mapping is due to the freedom in redefining the BF fields. This is the ultimate reason
for the presence of the free parameters ($B_1$, $B_2$...) in the mapping  seen previously. In fact the exact
inverse mapping is given by

\begin{eqnarray}\label{mapping inverso}
\widehat A_\mu&=&\left(\widehat C_{AA}P_{\mu,\nu}+\delta_{\mu\nu}\right) A^\nu +\widehat C_{AB} S_{\mu\alpha\beta} B^{\alpha\beta},\nonumber\\
\widehat B_{\mu\nu}&=&\widehat C_{BA}S_{\mu\nu\alpha} A^\nu +\left(\widehat C_{BB} P_{\mu\nu,\alpha\beta}+\frac12\delta{\mu\nu,\alpha\beta}\right)
 B^{\alpha\beta},
 \end{eqnarray}
 
 \noindent with
 \begin{eqnarray}\label{strucfunc2}
\widehat C_{AA}&=&\frac{{\sigma}}{2^{\frac32}m^{\frac12}\Box }\left[\frac{(
m-\sqrt{m^2+\Box})^{\frac32}}{(m^2+\Box)^{\frac12}}\right]  -\frac1\Box ,\nonumber\\
\widehat C_{BB}&=&\frac{{2}^{\frac12}}{\sigma m^{\frac12}\Box }\left[\frac{(
m-\sqrt{m^2+\Box})^\frac 32}{(m^2+\Box)^\frac12}\right] -\frac1\Box ,\nonumber\\
\widehat C_{AB}&=&\frac{{\sigma}}{2^{\frac52}m^{\frac12}}\left[\frac{
\left(m-\sqrt{m^2+\Box}\right)}{\left(m^2+\Box\right)}\right]^{\frac12} , \nonumber\\
\widehat C_{BA}&=&\frac{{2}^{\frac12}}{\sigma m^{\frac12}}\left[\frac{
\left(m-\sqrt{m^2+\Box}\right)}{\left(m^2+\Box\right)}\right]^{\frac12}.  
\end{eqnarray}

The iterative mapping may be retrieved by expanding  the structure functions in terms of $\frac{\Box}{m^2}$ and, at the same time, expressing 
the operator ${\sigma}$ in terms of arbitrary parameters as
\begin{equation}
{\sigma}=\sum_{n=0}^\infty C_n\left({\frac{\Box}{m^2}}\right)^n
\end{equation}
and similarly to its inverse. With this equation (\ref{mapping inverso}) will reproduce 
 equations
(\ref{bff6a}).
The independent parameters $B_j$ are thus seen to owe their origin to the freedom in defining the
operator $\sigma$. This allows for an independent parameter, $C_j$, to be introduced 
 at each order in $\Box^j$.

\section{Dimensional reduction considerations}

Let us consider in 5-D the model with the antisymmetric field which represents
a direct generalization of the Maxwell-Chern-Simons model

\begin{equation}\label{mcs}
S=\int d^5x\left( \frac 1{12}H_{\mu \nu
\rho }H^{\mu \nu \rho }+\frac m{12}\varepsilon ^{\mu \nu \alpha\beta\rho }B_{\mu
\nu }H_{\alpha\beta\rho }\right) \,.  \label{bffgh3}
\end{equation}

The mapping of the field to the pure Chern-Simons field

\begin{equation}\label{csi}
\widehat S=\int d^5x\left(\frac m4\varepsilon ^{\mu \nu \alpha \beta \rho}B_{\mu
\nu }H_{\alpha\beta\rho}\right) \,  \label{bffgh4}
\end{equation}
is implemented by the transformation

\begin{eqnarray}\label{transform5d}
 B_{\mu\nu}&=& \hat B_{\mu\nu}+
 \left[\frac{{2}^{\frac12}}{2 m^{\frac12}\Box }\left[\frac{(
m-\sqrt{m^2+\Box})^\frac 32}{(m^2+\Box)^\frac12}\right] -\frac1\Box \right]
 P_{\mu\nu ,\alpha\beta}\widehat B^{\alpha\beta}\nonumber\\
 &&\hskip 1cm -\frac{{2}^{\frac12}}{12 m^{\frac12}}\left[\frac{
\left(m-\sqrt{m^2+\Box}\right)}{\left(m^2+\Box\right)}\right]^{\frac12}  
\varepsilon_{\mu\nu\alpha\beta\rho}\widehat H^{\alpha\beta\rho}.
\end{eqnarray}
 This result is obtained repeating the argument of section 2-b in the 5-dimensional
 spacetime. Note that in this case the topological model does not present any freedom in
 rescaling the fields as occurs in 4D. 
  The dimensional reduction of the model (\ref{mcs}) by precluding
 any dependence on the variable $x^4$ so that $B_{\mu\, 4}:=A_\mu$ 
  leads to the Cremmer-Sherk model (\ref{cs}). Under similar considerations
   the topological model
  (\ref{csi}) is led to the pure BF model (\ref{purebf}). Within this setting the 
 mapping of the 5D fields (\ref{transform5d}) is reduced to (\ref{mapping1}) and 
 (\ref{strucfunc}) if we eliminate the freedom in the mapping by identifying $\sigma=2$.
 Thus the dimensional reduction turns out to give a criterion to fix in a natural fashion the
 mapping of the fields.

As we have seen, the mapping connecting the model with topological mass
generation in four dimensions to the pure topological $BF$ model is related
to similar properties of models in 5-dimensional spacetime which present only the 
antisymmetric field. In this
section we perform one more step in the dimensional reduction program presenting the 
similar property appearing in the model obtained
after the dimensional reduction of the Cremmer-Sherk's action to 3D. The reduced
action is given by 
\begin{eqnarray}\label{bff111}
\mathcal{S} &=&\mathcal{S}_{top}+\mathcal{S}_{ntop}  \nonumber  \\
&=&\int d^3x\left( -\frac m6\varepsilon ^{\mu \nu \rho }\varphi H_{\mu \nu
\rho }+\frac m2\varepsilon ^{\mu \nu \rho }c_\mu F_{\nu \rho }\right) + \nonumber\\
&&\int d^3x\left( -\frac 14G^{\mu \nu }G_{\mu \nu }-\frac 14F^{\mu \nu
}F_{\mu \nu }+\frac 1{12}H_{\mu \nu \rho }H^{\mu \nu \rho }+\frac 12\partial
^\mu \varphi \partial _\mu \varphi \right)  \nonumber
\end{eqnarray}
where, after the reduction, 
\begin{eqnarray*}
A_\mu &\rightarrow &A_{\mu ,}\,\,\,\varphi \,, \\
B_{\mu \nu } &\rightarrow &B_{\mu \nu },\,\,C_\mu \,, \\
G_{\mu \nu } &=&\partial _\mu C_\nu -\partial _\nu C_\mu \,.
\end{eqnarray*}
The mapping to ensure that
\[
\mathcal{S}_{top}\left( \widehat{A}_\mu ,\widehat{B}_{\mu \nu },\,\widehat{%
\varphi },\,\widehat{c}_\mu \,\right) =\mathcal{S}_{top}\left( A_\mu ,B_{\mu
\nu },\varphi ,c_\mu \right) +\mathcal{S}_{ntop}\left( A_\mu ,B_{\mu \nu
},\varphi ,c_\mu \right) , 
\]
will be given by

\begin{eqnarray}
\widehat{A}_\mu &=&A_\mu +\sum_{n=1}^\infty \frac 1{{(2m)}^n}\vartheta _\mu ^n\,,
\\
\widehat{B}_{\mu \nu } &=&B_{\mu \nu }+\sum_{n=1}^\infty \frac 1{{(2m)}^n}\phi
_{\mu \nu }^n\,, \\
\widehat{\varphi }\, &=&\varphi +\sum_{n=1}^\infty \frac 1{{(2m)}^n}\alpha ^n\, \\
\widehat{c}_\mu &=&c_\mu +\sum_{n=1}^\infty \frac 1{{(2m)}^n}k_\mu ^n\,.
\end{eqnarray}
Following the same lines as in the previous section the coefficients are iteratively
defined. The first ones are given by 

\begin{eqnarray}
\vartheta _\mu ^1 &=&-\frac 12\varepsilon _{\mu \nu \rho }G^{\nu \rho }\,, \\
\vartheta _\mu ^2 &=&(1-B_1)\partial ^\nu F_{\nu \mu }\,,  \nonumber \\
\vartheta _\mu ^3 &=&\frac {B_1}{2}\varepsilon _{\mu \alpha \beta }\Box G^{\alpha
\beta }\,,  \nonumber \\
\vartheta _\mu ^4 &=&(-1-B_1+B_1^2)(1-B_2)\Box\partial ^\nu F_{\nu \mu }\,,   \\
&& \nonumber\\
\phi _{\mu \nu }^1 &=&-\varepsilon _{\mu \nu \rho }\partial ^\rho
\varphi \,,  \nonumber \\
\phi _{\mu \nu }^2 &=&\frac{B_1}{3}\partial ^\alpha H_{\alpha \mu \nu }\,,  \nonumber \\
\phi _{\mu \nu }^3 &=&-(1-B_1)\varepsilon _{\mu \nu \rho }\Box
\partial ^\rho \varphi \,,  \nonumber \\
\phi _{\mu \nu }^4 &=&(-1-B_1+B_1^2)B_2\Box\partial ^\alpha H_{\alpha \mu \nu }\,, 
\end{eqnarray}

\begin{eqnarray}
\alpha ^1 &=&-\frac 1{6}\varepsilon _{\mu \nu \rho }H^{\mu \nu \rho } \nonumber\\
\alpha ^2 &=&(1-B_1)\Box\varphi \nonumber\\
\alpha ^3 &=&\frac {B_1}{6}\varepsilon _{\mu \nu \rho }\Box H^{\mu \nu
\rho } \nonumber\\
\alpha ^4 &=&(-1-B_1+B_1^2)(1-B_2)\Box ^2\varphi \\
&& \nonumber\\
\beta_\mu ^1 &=&\frac 12\varepsilon _{\mu \nu \rho }F^{\nu \rho } \nonumber\\
\beta_\mu ^2 &=&B_1\partial ^\nu G_{\nu \mu } \nonumber\\
\beta_\mu ^3 &=&\frac {1-B_1}{2}\varepsilon _{\mu \nu \rho }\Box F^{\nu \rho }\nonumber \\
\beta_\mu ^4 &=&B_2(-1-B_1+B_1^2)\Box\partial ^\alpha G_{\alpha \mu } .
\end{eqnarray}

It should be clear that the above expression may be summed up leading to expressions that
parallel the exact mapping obtained in D=4. Indeed the classical argument of mapping the
actions as above depicted allows to obtain the coefficients in dimensionally reduced models
from the knowledge of the 
higher dimensional mapping coefficients. Since the propagators are the inverses of the operators defining the actions their exact mapping in lower dimensions can also be read from
the corresponding expressions in higher dimensions. The non-local operators that map
the local fields are essentially the same ones.

\subsection{ Cohomological argument}

With the aim of giving a cohomological argument to equations $\left( \ref
{bff1}\right) $ and $\left( \ref{bff3}\right) $ we use the
Batalin-Vilkovisky formalism. Five antifields are required $(A^{*\mu
},B_{\mu \nu }^{*},c^{*},\eta ^{*\mu },\rho ^{*}),$ corresponding
respectively to the gauge connection $A_\mu $, to the antisymmetric tensorial
field $B_{\mu \nu }$, the ghost of Faddev-Popov $c$, the ghost which comes
from the gauge transformation of $B_{\mu \nu }$ and to the ghost $\rho $.
The last ghost appears because of the reducibility of the $B_{\mu \nu }$
symmetry, which requires, to the complete fixation, one extra ghost.

As mentioned before, the cohomological analysis of the BRST differential
operator identifies the terms which can, or cannot, be reabsorbed by field
redefinitions $\cite{ng3}$. In this context, quantities which do not belong
to the cohomology of this operator can be, at the classical level, absorved
by field redefinitions. As we will show now, terms such as $F_{\mu \nu
}F^{\mu \nu }$ and $H_{\mu \nu \rho }H^{\mu \nu \rho },$ contained in the
action $\left( \ref{bff1}\right) ,$ are not out of this rule.

To see the triviality of these terms lets write the actions as 
\[
\Sigma =S+\Sigma _{ant}, 
\]
where 
\[
\Sigma _{ant}=\int d^4x\left( A_\mu ^{*}\partial ^\mu c+B_{\mu \nu
}^{*}\partial ^\mu \eta ^\nu +\eta _\mu ^{*}\partial ^\mu \rho \right) . 
\]
The nilpotent BRST operator is given by 
\begin{eqnarray*}
s &=&\int d^4x\left( \frac{\delta \Sigma }{\delta A_\mu ^{*}}\frac \delta
{\delta A^\mu }+\frac{\delta \Sigma }{\delta A^\mu }\frac \delta {\delta
A_\mu ^{*}}+\frac{\delta \Sigma }{\delta B_{\mu \alpha }^{*}}\frac \delta
{\delta B^{\mu \alpha }}+\frac{\delta \Sigma }{\delta B^{\mu \alpha }}\frac
\delta {\delta B_{\mu \alpha }^{*}}\right. \\
&&\left. \frac{\delta \Sigma }{\delta c^{*}}\frac \delta {\delta c}+\frac{%
\delta \Sigma }{\delta c}\frac \delta {\delta c^{*}}+\frac{\delta \Sigma }{%
\delta \eta _\mu ^{*}}\frac \delta {\delta \eta ^\mu }+\frac{\delta \Sigma }{%
\delta \eta ^\mu }\frac \delta {\delta \eta _\mu ^{*}}+\frac{\delta \Sigma }{%
\delta \rho ^{*}}\frac \delta {\delta \rho }+\frac{\delta \Sigma }{\delta
\rho }\frac \delta {\delta \rho ^{*}}\right) \,.
\end{eqnarray*}
The action of this operator on the fields and anti-fields is given by: 
\begin{eqnarray}
sA_\mu &=&\partial _\mu c  \label{bff6} \\
sB_{\mu \nu } &=&\partial _\mu \eta _\nu -\partial _\nu \eta _\mu  \nonumber
\\
s\eta _\mu &=&\partial _\mu \rho  \nonumber \\
sc &=&s\rho =0  \nonumber \\
sA_\mu ^{*} &=&\partial ^\nu F_{\nu \mu }+\frac m6\varepsilon _{\mu \nu \rho
\sigma }H^{\nu \rho \sigma }  \nonumber \\
sB_{\mu \nu }^{*} &=&-\partial ^\rho H_{\rho \mu \nu }+\frac m2\varepsilon
_{\mu \nu \rho \sigma }F^{\rho \sigma }\,,  \nonumber\\
s\eta _\mu ^{*} &=&\partial ^\nu B_{\mu \nu }  \nonumber \\
sc^{*} &=&\partial ^\mu A_\mu ^{*}  \nonumber \\
s\rho ^{*} &=&\partial ^\mu \eta _\mu ^{*}  \nonumber
\end{eqnarray}

\noindent The triviality of these terms is strongly based in the BRST
transformations form of the antifields $A^{*a\mu }$ and $B_{\mu \nu }^{*a}$,
given by $\left( \ref{bff6}\right) .$ Contracting conveniently the
expressions contained in $\left( \ref{bff6}\right) $ with the $\varepsilon $
- tensorial density 
 we obtain 

\begin{equation}
H_{\mu \nu \rho }=\frac 1m\left( \varepsilon _{\mu \nu \rho \sigma
}sA^{*\sigma }-\varepsilon _{\mu \nu \rho \sigma }\partial _\chi F^{\chi
\sigma }\right) \,.  \label{bff7b}
\end{equation}

and 

\begin{equation}
F_{\mu \nu }=-\frac 1{2m}\left( \varepsilon _{\mu \nu \rho \sigma }sB^{*\rho
\sigma }+\varepsilon _{\mu \nu \rho \sigma }\partial _\chi H^{\chi \rho
\sigma }\right)   \label{bff7a}
\end{equation}

The equations $\left( \ref{bff7b}\right) $ and $\left( \ref{bff7a}\right) $
can be used in a recursive way, \ 
\begin{equation}
H_{\mu \nu \rho }=\frac 1ms\sum_{n=0}^\infty \left( \frac \partial m\right)
^{2n}\left( \varepsilon _{\sigma \mu \nu \rho }A^{*\sigma }-\frac 1mH_{\mu
\nu \rho }^{*}\right) \,,  \label{dgfr}
\end{equation}

\begin{equation}
F^{\mu \nu }=-\frac 1{2m}\varepsilon ^{\mu \nu \rho \sigma }s\left\{ B_{\rho
\sigma }^{*}+\frac 1m\sum_{n=0}^\infty \left( \frac \partial m\right)
^{2n}\left( \frac 12\varepsilon _{\alpha \eta \rho \sigma }F^{*\alpha \eta
}-\frac 1m\partial ^\alpha H_{\alpha \rho \sigma }^{*}\right) \right\} 
\label{fgd}
\end{equation}
where 
\[
F_{\mu \nu }^{*}=\partial _\mu A_\nu ^{*}-\partial _\nu A_\mu ^{*}\,\,\,\,\, 
\]
and 
\[
H_{\mu \nu \rho }^{*}=\partial _\mu B_{\nu \rho }^{*}+\partial _\rho B_{\mu
\nu }^{*}+\partial _\nu B_{\rho \mu }^{*}\,.\, 
\]
Since $sH_{\mu \nu \rho }=sF_{\mu \nu }=0,$ the triviality of the curvatures 
$F_{\mu \nu }=s\Lambda _{\mu \nu }$ and $H_{\mu \nu \rho }=s\Gamma _{\mu \nu
\rho }$, which clearly appears in the equations $\left( \ref{dgfr}\right) $
and $\left( \ref{fgd}\right) ,$ can be used to show that the terms $\int F^2$
and $\int H^2$ take the final form

\[
\int d^4xF^{\mu \nu }F_{\mu \nu }=s\int d^4xF^{\mu \nu }\Lambda _{\mu \nu } 
\]
and 
\begin{equation}
\int d^4x\left( H_{\mu \nu \rho }H^{\mu \nu \rho }\right) =s\int d^4xH^{\mu
\nu \rho }\Gamma _{\mu \nu \rho }  \nonumber
\end{equation}
which lead us to conclude that they can be reabsorbed by redefinitions of
fields.

The above analysis is similar to that presented in reference $
\cite{setup}$. Looking at the BRST transformations of the antifields $A^{*}$
and $B^{*},$ given by $\left( \ref{bff6}\right) ,$ we see that the terms
which lead to the conclusion of the triviality of the curvatures are $
\varepsilon _{\mu \nu \rho \sigma }H^{\nu \rho \sigma }$ and $\varepsilon
_{\mu \nu \rho \sigma }F^{\rho \sigma }.$ Clearly, these terms are related
to the topological term $\varepsilon BF.$ We see then that, in the cases of
Chern-Simons and Cremmer-Sherk, the presence of the topological terms leads
to the existence of field iterative redefinitions that we have presented.

\section{Conclusions}

In this work we have studied  in some detail a generalization from 2+1D
to 3+1D of a procedure that was known to map the MCS field to CS
field. In (3+1)D the Cremmmer-Sherk model is mapped to the Abelian
version of the BF model. This mapping has been established both within
an iterative as well as within an exact procedure. One remarkable new aspect 
that emerges is the presence of a great deal of freedom in the
mapping in four dimensions. This freedom has been elucidated as
due to the form of the pure topological action which is defined
through mixed products of fields. The  invariance under rescaling of 
the fields of the
BF type action is responsible for it. Since this kind
of action is naturally considered in even dimensional topological models,
 the non-uniqueness in the mapping should be expected to hold in even dimensions.

The knowledge of the exact mapping provides us with a typical scale,
given by the mass parameter $m$. The mapping may be used for instance
for computing loop variables of the Cremmer-Sherk model using the corresponding
expressions of the pure BF model. This suggests to perform the
computation in closed fashion without resource to expansions given
by the iterative mapping.  In any case the mass parameter $m$ may provide
valuable hints to discern in which cases computations using the iterative
mapping should or not be considered reliable. It can even provide alternative
expansions for instance in direct powers of $m$ instead of the inverse 
power series provided by the iterative mapping. 

The dimensional reduction arguments here presented relates the mechanism of
mapping from more involved actions to structurally minimized models in 
different dimensions. 
Besides providing a criterion to fix the mapping, the kinematical
 dimensional reduction may offer
insights as to the low momenta field variables needed in dimensional reductions 
of high temperature limits in field theory. Actions of the Cremmer-Sherk type
are expected to play a role in approaches where current fermions condensates are 
explicitly controlled with a bosonization scheme.
The high temperature limit of QED under this setting will lead to a dimensional
reduction paralleling the one here provided.

The cohomological argument here presented should be understood as giving
consistency to the mapping among the fields in the iterative approach.
Clearly this iterative mapping prevents one from considering the 
limit $m\longrightarrow 0$
for which the coefficients would become singular.
Nevertheless the explicit knowledge of the exact mapping allows one to consider this
limit after summing the series. Indeed  this limit in equations 
(\ref{strucfunc}) may be performed smoothly. One should  first 
consider the mass parameter $m$ of the pure BF model as independent from
the corresponding parameter  in the Cremmer-Sherk model. The structure constants 
become simpler non-local
functions since they do not present any mass scale in their definitions.

In order to  properly appreciate the physical meaning of the mapping, it
is important 
 to call attention to the necessity of defining the physical content
of a local field theory in terms of the local polynomial algebra of 
observable fields. The mapping here provided relates two local models each with
its physical Hilbert space reconstructed from the Wightmann functions of its own polynomial
algebra\cite{strocchi,belvedereerubens}. Since the mapping involves  non-local functions
it should be clear that within the pure BF model there are two Hilbert spaces to be obtained.
One Hilbert space is obtained from the  local polynomial algebra of
fields defined after expressing the Cremmer-Sherk fields non-locally in terms of the pure BF model
fields and it should not be confused
with the Hilbert space of the pure BF model itself. This later is obtained from its local polynomial 
algebra of fields. Although constructed with the same model fields the first Hilbert space is not  isomorphic to the second one. Instead, it will be isomorphic to the
 Hilbert space of the Cremmer-Sherk model. The same reasoning goes in the other
 direction of the mapping. In this context it is clear that neither
 Hilbert space should be considered as a subspace of the other. It is not a mapping of
 physical states that is being addressed here but a non-local mapping among the fields.
 
  The generality of the mapping here considered can be further enhanced by introducing
arbitrary scalar operators in the definitions of the quadratic non mixed
 terms of the the vector and antisymmetric fields and considering the parameter
 $m$ as an scalar operator acting either on vector or antisymmetric field.
 This generalized gauge invariant action will be mapped to the pure BF model
 in a very similar  way with the structure functions of the mapping being 
 slightly modified. Furthermore it is to be expected\cite{setup} that the
 introduction of arbitrary gauge invariant interaction terms can be absorbed
 by considering nonlinear mappings.

\section{Acknowledgements}

We are indebted to N. A. Lemos for carefully reading this manuscript.

\end{document}